# High School Students' Group Interaction with the Electric Field Hockey Interactive Simulation

Ying Cao

Abstract: This study investigates a group of high school students in a physics classroom interacting with a computer simulation that simulates the electrostatic interaction as a hockey game, the Electric Field Hockey. The activity featured in this study took place prior to the students receiving formal instruction about the electric field. The learning goal was to allow students to explore the simulated electrostatic phenomena. The study asks the following research question: How do high school students collectively explore simulated electrostatic phenomena while interacting with the Electric Field Hockey computer simulation during a classroom activity? In this paper, through careful analysis of classroom videos, I present a case study about the group mentioned above. The results show that high school students encountered and dealt with multiple types of tensions in the activity and explored the electrostatic phenomena simulated by the Electric Field Hockey. Students prioritized exploring different types of phenomena when they encountered different types of tensions. Based on the findings I propose: When carrying out an activity with a novel educational tool, a teacher can attend to, and jump in at appropriate points, when tensions arises thus to orient students to explore the simulated physical phenomena in more productive ways.

The concept of the electric field is usually introduced in high school physics and continues to be addressed in college courses. The concept is key to physics theories and widely applied to science and engineering (e.g., electric engineering). Therefore, learning of the electric field has been emphasized in physics curricula at both the high school and college levels. However, physics education research studies have found that students have difficulties understanding the theories (Maloney, O'Kuma, Hieggelke, & Van Heuvelen, 2001; Rozier & Viennot, 1991), concepts (Furio & Guisasola, 1998), and/or representations (Tornkvist, Pettersson, & Transtromer, 1993) related to the electric field. In other words, previous studies have indicated that students often fail to understand the subject matter (see a review in Chapter 3 of this dissertation).

Among many possible reasons for students' seemingly unsuccessful learning of the electric field, I highlight one as the focus of the present study: the lack of experience with the electrostatic[1] phenomena that they are to learn. Scientific inquiry begins with experiencing the real world. Based on their experience, scientists generalize laws of physical phenomena. In school, most science lessons open with an experimental demonstration followed by the theoretical explanation of the phenomena shown in the demonstration. Most electrostatics lessons start with demonstrating (with lab equipment) Coulomb's law, which generalizes the law of interaction between two point charges (an ideal model of individual charges). However, to demonstrate the related phenomena is not easy. In everyday life students' experiences with electricity—e.g., thunder and lightning, electric circuits—involve the behavior of numerous electric charges. These phenomena do not evidently demonstrate Coulomb's law, which describes the interaction between two point charges. Therefore, students do not have much previous experiences that they can relate to Coulomb's law. An alternative way to demonstrate Coulomb's law is to rely on lab equipment that is designed to show the interaction between point charges (such as an electroscope). Unfortunately, it is still easy to fail at demonstrating the phenomena because the effectiveness of the demonstration is vulnerable to humidity (the electric charges will "vanish" when exposed to humid air). Overall, demonstrating electrostatic phenomena is notoriously hard.

---

[1] In this paper, I use the terms electric field (as a physical theory) and electrostatics interchangeably. I also use the terms electric field (as a physical entity) and electrostatic phenomena interchangeably.

Learning physics is not memorizing formulas without making sense of the connection between a formula and the related phenomenon. Therefore, having some experience with the phenomenon is indispensible for the learning of electrostatics. The challenge of demonstrating electrostatic phenomena in the real world is somewhat unconquerable. As another alternative, some physics education researchers have suggested that if applied appropriately, a computer simulation can substitute the real-world laboratory to demonstrate the physical phenomena that students are going to learn (Finkelstein et al., 2005). The Electric Field Hockey (E-Field Hockey) game (http://phet.colorado.edu/en/simulation/electric-hockey) is such an interactive simulation developed by PhET, a University of Colorado educational technology group that designs interactive simulations for educational purposes. The E-Field Hockey is the focus of the present study, which I will describe in detail in later sections.

Other than the E-Field Hockey, some researchers and practitioners have designed computer-simulated 3-D visualizations of electromagnetic field lines to help college students learn electromagnetism (e.g., Barnett, Grant, & Higginbotham, 2004; Belcher & Olbert, 2003; Dori & Belcher, 2005). Dori and Belcher's findings have shown the positive impact that 3-D visualizations have had on students' answers to assessment questions, which indicates the advantage of including computer-simulated visualizations in lessons that address abstract physics concepts such as the electric field.

The E-Field Hockey is another example that provides simulated electrostatic phenomena for educational use. It simulates the electric charges and interactions and allows a player to operate on the simulated charges through a computer mouse (e.g., drag and move electric charges on the electric field) and see the impacts (the consequent moving trajectories of the test charge). The literature has not reported the use of Electric Field Hockey interactive simulation. The present study presents a case study focused on a group of high school students as they were engaged in a classroom activity interacting with the E-Field Hockey as a group and explored the simulated electrostatic phenomena.

## The Electric Field Hockey

The E-Field Hockey interactive simulation provides students opportunities to experience simulated electrostatic phenomena. The Electric Field Hockey simulates individual electric charges and their interaction according to electrostatic theories (e.g., Coulomb's law) and Newtonian kinetics. It invites students to interact with the simulation by choosing the type and number of source charges and deciding the spots to put them on the simulated hockey field. The Electric Field Hockey allows students to observe the impact of source charges on a test charge in the latter's moving trajectory. In this section, I will describe some features of the E-Field Hockey relevant to the present study.

As shown in the screenshots (see Figure 1), on the Electric Field Hockey interface, electrically charged objects are depicted as small balls that appear in one of two colors: red balls represent positive charges; blue balls, negative charges. The balls are set up to be stored separately, according to their charge types, in one of the two storage bins in the top-right corner of the screen. The balls can be dragged by the mouse cursor, one at a time, and placed at different spots in a simulated square-shaped field. The black hockey puck that is set at rest on the left side represents a positive test charge. The location of the puck in Figure 1 is the puck's starting point. The object is to make the hockey puck hit the goal on the right side (the blue half-square bracket). To interact with the E-Field Hockey, a person starts by dragging and placing the red balls and blue balls anywhere on the field based on his or her prediction as to whether the red balls and blue balls that interact with the hockey puck can make it hit the goal. The person then hits the start button at the bottom left of the interface. The hockey puck moves according to the attraction and repulsion of the balls that surround it. If the hockey puck arrives at the goal, the computer emits a cheerful sound, *Dah dah…,* and the word *goal* is displayed on the screen in a large font size. If the hockey puck hits any barriers or rushes off the field, then the person does not achieve the goal of this round. The game has four challenge levels: a practice level (Figure 1, top left) and levels 1 to 3 (Figure 1, top right, bottom left, and bottom right, respectively). The practice level has no walls obstructing the puck's path between the starting position and the goal. Levels 1 to 3 have differing numbers and patterns of walls. When playing with these three levels, the person must make the hockey puck move in a way that it can get around the wall(s) in order to hit the goal.

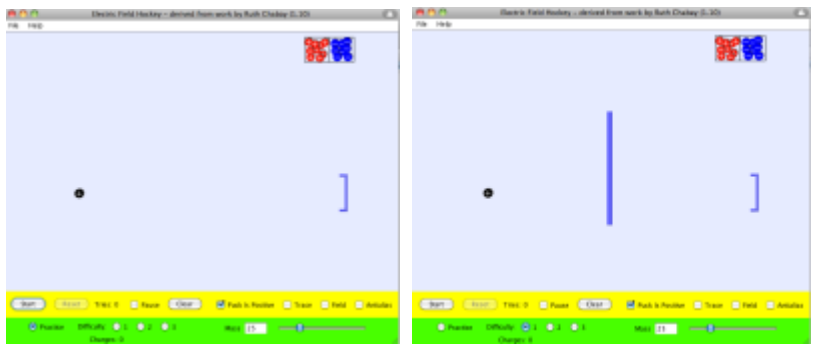

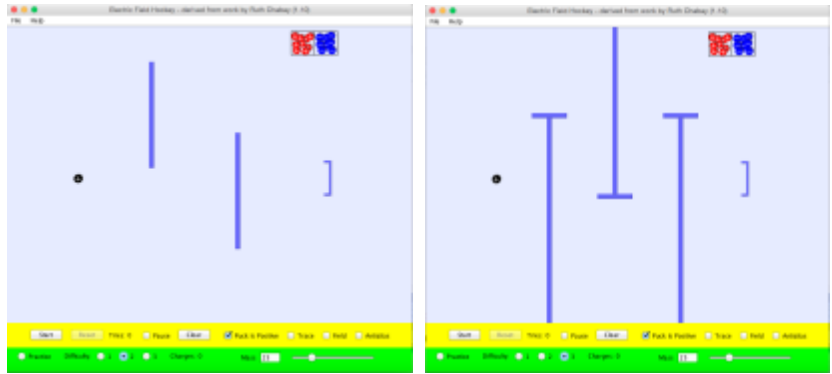

Figure 1. E-Field Hockey interactive simulation interface. Top left: Practice level; top right: Level 1; bottom left: Level 2; bottom right: Level 3.

At the bottom of the interface, there is a yellow menu bar atop a green one (see Figure 2). The start button is on the left end of the yellow menu bar. Next to the Start button, there is a reset button that returns the hockey puck to its starting point. The yellow menu bar also displays the number of tries a player has attempted, a pause check box to stop the action, and a clear button to eliminate all the red balls and blue balls that have been placed in the field (so that the player can start over with a clean field). The hockey puck's charge type can be assigned by checking or unchecking a box labeled "Puck is Positive" in the middle of the yellow bar.

Three check boxes to the rightmost section of the yellow menu bar are labeled "Trace," "Field," and "Antialias." Checking the trace box makes the trajectory of the puck visible. When the puck moves, it leaves behind a trace of its path in a red, dashed line. Checking the field box shows the configuration of electric field lines in numerous short arrows spread over the hockey field. Checking the antialias box makes the field lines darker. Electric field lines represent the electric field vector at each point of the field. When one adds, subtracts, and moves the charged balls in the field, the field lines change accordingly in real time.

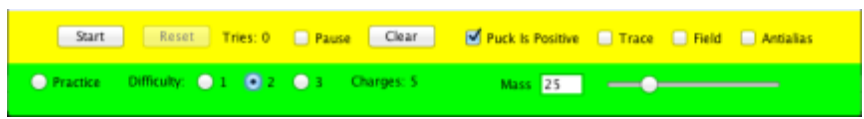

Figure 2. Menu bars at the bottom of the interface.

In the green menu bar at the bottom, one can choose difficulty levels: Practice and Levels 1 to 3. The green menu bar also shows the total number of electric charges (i.e., the total number of red and blue balls) that have been placed in the field. On the right end of the green menu bar, one can set the hockey puck's mass by manipulating the slider. The puck's mass is shown in the white box (the unit is not specified). I will discuss some of these features further when I present the data in later sections.

The E-Field Hockey interactive simulation depicts point charges (an ideal model of electrically charged objects) and simulates the interaction among the charges in a fun, friendly, and gamelike environment. It gives students the option to manipulate the source charges and see the impact on a test charge. It has gradient challenge levels that can motivate students to explore more about the simulated electrostatic phenomena. Implementing a class activity with the E-Filed Hockey, my research question in the present study is: In what ways the students as a group collectively explore the simulated electrostatic phenomena?

**Theoretical Framework**

I adopt Activity Theory (AT) as the theoretical and analytical framework to examine the E-Field Hockey activity. Rooted in the social-cultural tradition of learning (e.g., Vygotsky, 1978), AT claims that learning is a collective endeavor (Leont'ev, 1981) that involves multiple elements in multiple ways. AT is helpful as the analytical lens for the present study because the theory recognizes different, and different types of, elements within an activity system. It also acknowledges the complicated interactions within and among the elements. The theory helps me disentangle a complex activity (the E-Field Hockey classroom activity) into analyzable elements, and keep track of the progress of the activity by attending to the interactions within and among the elements.

Activity theory emphasizes the mediating role of a learning tool between the learner and the object. It also includes social elements beyond the individual level that views learning as a collective endeavor. The present study focuses on the role of the E-Field Hockey (the tool) in students' (the learner) exploration of electrostatic phenomena (the object) as a group in a high school classroom (the social context). AT can help me examine the activity through these elements and understand how students collectively made use of the E-Field Hockey to explore the simulated electrostatic phenomena in classroom.

**The Activity Model**

Figure 3 models the structure of an activity system according to Engeström (1987). The theory is developed from the socio-cultural tradition that emphasizes that learning is not a linear interaction between the *subject* and the *object*, but a mediated experience through cultural *tools* (such as materials and language). Vygostky represented this idea through a triangle (Vygotsky, 1978; see the top triangle in Figure 3). AT scholars built on Vygostky's work and argued that the three elements in Vygotsky's triangle—*subject*, *object*, and *tool*—constitute only the "tip of the iceberg" of an activity system. They expanded the theory to include more and specific cultural elements—*rules*, *community*, and *division of labor* (as shown at the bottom of the diagram in Figure 3).

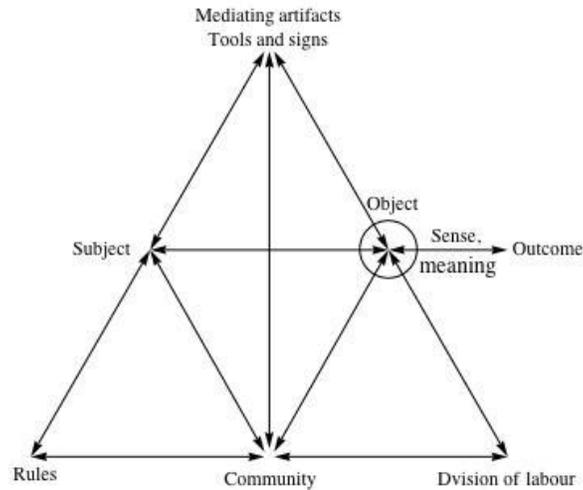

Figure 3. The structure of a human activity system (Engeström, 1987, p. 78).

In the AT model shown in Figure 3, the subject element refers to the learner. In the E-Field Hockey activity featured in the present study, the *subject* was the group of high school students participating in the E-Field Hockey activity. The *object* is the problem space, or the goal, of the activity. In the E-Field Hockey activity, the object was to explore simulated electrostatic phenomena. For students, their most apparent goal when interacting with the E-Field Hockey may be to make the puck hit the goal. In order to achieve this students will be exploring certain phenomena that they anticipate can show the hockey puck moving and arriving at the goal. There are other phenomena that can also been created and observed in the Electric Field Hockey but can have nothing to do with making the puck arrive at the goal (such as making the hockey puck moving along a circle far away from the goal, or just making the puck move randomly on the field). Whether students will also explore these latter phenomena remains a question for data analysis. The *mediating tool* in the present study was the E-Field Hockey computer simulation. The *rules* included, but were not limited to, classroom regulations, cultural norms, and the game's rules that had been coded by the developers in the software. The *community* consisted of the students and me in the classroom. The Electric Field Hockey was installed in the teacher's computer. During the E-Field Hockey activity the computer's screen was projected onto the front board of the classroom. The *division of labor* was that one student at a time operated the mouse and played the game, while other students gave suggestions to the player about how to play. Students volunteered to operate the Electric Field Hockey and took turns with the mouse. The teacher began the activity with an introduction and then let students operate the mouse, assisting from the side to circulate the mouse and answering students' questions at times.

According to AT, all the elements are inter-related to form an activity system (represented by the arrows in Figure 3 connecting all the elements in the system). The system is object-oriented (represented by the circle around the object element, in Figure 3). Through the interactions among the elements, the activity system is transformed to exhibit some outcome (represented by the branch out from the object element on the right side of the triangular model in Figure 3). In the case of the present study, the activity was oriented by the object, exploring simulated electrostatic phenomena. Students' gained experience with the simulated electrostatic phenomena (such as the attraction and repulsion of electric charges; the impact of electric interactions on a test charge's moving trajectory, etc.) were one outcome of this activity.

**The Principle of Contradictions**

AT has five principles to understand an activity system: (1) the activity system is the primary unit of analysis; (2) the multi-voicedness of the activity system; (3) the historicity of the activity system; (4) the central role of contradictions as sources of system change and development; and (5) the possibility of expansive transformations in the activity system (Engeström, 2001). The fourth principle about the role of contradictions focuses on the mechanism of system change, viewing contradictions as the driving force of change and development in activity systems (Il'enkov, 1977, 1982). Contradictions, according to AT scholars, are disruptions (Berge & Fjuk, 2006), "problems, ruptures, breakdowns, clashes" in activities (Kuutti, 1996, p. 34), and learning happens when contradictions are resolved (Murphy & Rodriguez-Manzanares, 2008). The principle of contradictions has been adopted by many empirical researchers studying educational technologies (see a review by Murphy & Rodriguez-Manzanares, 2008).

I apply this principle to examine the process through which students interact with the E-Field Hockey activity because AT claims that contradictions are the driving forces of change and development of the activity. For the present study, I describe contradictions as tensions (also used in Basharina, 2007; Berge & Fjuk, 2006; and Engeström, 2001) that involve two and more competing factors in the activity. During the E-Field Hockey activity featured in this study, there was no explicit leading or instruction from the teacher. Students explored all the levels of the E-Field Hockey. They won the easier levels and moved on with the higher levels. Without instruction, students also explored more about the E-Field Hockey tool itself (such as the embedded representation of field lines and force lines). Given that AT claims tensions are inherent in any human activities and the tensions drive an activity system change, I anticipate some tensions in the E-Field Hockey activity and speculate that the tensions should play a role while students as a group work toward their goal(s). I therefore narrow down my research question in two more specific questions:

(1) What were some tensions in the Electric Field Hockey activity?

(2) How did the tensions contribute to students' exploration of simulated electrostatic phenomena?

## Methods

**Context and Participants**

I implemented the E-Field Hockey activity with a group of 36 ninth-grade students enrolled in a summer program at a private school in China. The summer program provides courses focused on tenth-grade subjects (math, physics, chemistry, etc.). In these courses the teachers cover the first few units of the course that will be officially taught in high school the following fall. The idea of the summer program is to prepare students with new content and methods that they are going to learn in the next grade level of high school. Guided by this big idea, teachers in the program are granted freedom to develop their own syllabi according to their specific teaching goals and/or the students' skill levels. At the time the students participated in my study, most of them were between 15 and 16 years old and had not yet received formal instruction on electrostatics.

The physics course I taught consisted of 12 daily sessions spread during three weeks. Each session was 100 minutes long with a 10-minute break between the two halves of the session. During the 12-session course, I covered a unit on algebra-based, one-dimensional motion; a unit on the concept of force; and a unit on electricity, which is the focus of the present study.

The electricity unit was taught during the last five sessions of the course, in parallel with the mechanics units. Time allotted for the electricity unit on each day varied and depended on the time needed for the mechanics content. The electricity content took the remaining time of the session. I called the electricity part of the lesson "electricity time" and told the students that this was part of my research project. The electricity unit had two open-ended activities. The first activity was the E-Field Hockey activity featured in the present study. The second activity was the comic strip activity that has been described in previous chapters of this dissertation.

**The E-Field Hockey Activity**

The E-Field Hockey activity was implemented during the first three days of the five-day electricity unit. The game was installed on my computer and the computer's screen was projected on the front board during the activity. The whole class shared this computer during the activity. Students took turns with a wireless mouse to operate the E-Field Hockey. While one played, others gave suggestions. I introduced the game at the beginning of the activity, and then held back to let the students play. Students explored the E-Field Hockey for 22 minutes on Day 1 on the practice level (4 minutes) and level 1 (18 minutes), 10 minutes on Day 2 on level 2, and 17 minutes on Day 3 on level 3 (for a total of 49 minutes).

**Data Collection**

All three days of the activity were videotaped by multiple cameras: one camera at the back of the classroom recording the front board, and the other two cameras recording the students from the two corners of the front of the classroom. During the second and third days, I also recorded the screen of the computer on which the students played the game (due to a technical problem the first day's screen was not successfully recorded).

**Data Analysis**

In order to answer my research questions, the analysis sought to identify tensions in the Electric Field Hockey activity and how the tensions may have influenced students' exploration of the simulated electrostatic phenomena. According to AT, tensions are manifested through disturbances (Capper & Williams, 2004), which are "unintentional deviations from the script [which] cause discoordinations in interaction" and "deviations in the observable flow of interaction" (Engeström, Brown, Christopher, & Gregory, 1991, p. 91). In other words, disturbances are visible evidence of tensions. Therefore, in my analysis, I looked for disturbances in the classroom videos as manifestations of tensions. When a disturbance occurs, the activity flow is disturbed by someone or some phenomena that was not expected by most of the students. Therefore, there are usually visible signs of and/or reactions to that disturbance, such as students' exclamations, "Wow!" "Oops!" "Sigh…" and/or an explicit argument among students (e.g., while a student plays, some students suggest adding a blue ball in the upper right corner of the field; other students disagree and suggest not putting a ball there).

I examined the classroom transcript, marking all disturbances. Signs of disturbances included, but were not limited to, students' verbal expressions (e.g., individuals asking abrupt questions not in line with what most of the students were paying attention to, students' exclamations at certain findings), actions (e.g., clapping to express excitement), and intonations and/or facial expressions (e.g., showing encourage or discouragement). I then examined the vignettes each disturbance was nested in and interpreted the tensions manifested by each disturbance.

According to the principle of contradictions of Activity Theory, contradictions (or tensions) in the activity are related to the activity elements—subject, object, tool, rules, community, and division of labor (Engeström, 2001). Therefore, I tried to identify tensions according to these activity elements. For this purpose, I examined all the disturbances that I had marked and grouped them according to the activity elements that they were related to.

First, I grouped the disturbances related to the object element—to explore the electrostatic phenomena demonstrated in the E-Field Hockey. These disturbances were about the demonstrated physical phenomena shown on the screen. When students have developed a strategy—placing some positive and/or negative balls on the field and anticipating them will make the hockey puck move toward the goal and then hit the goal—the hockey puck's actual movements, very often, were out of students' expectation. Students, at these instances, often produced sounds such as "Ugh!" Wow!" or "Oops!" when they saw that the hockey puck did not go along the path that they had expected. These disturbances reflected the tension between students expectation of how the charges would behave and how the charges actually moved. For example, some disturbances occurred when students saw the hockey puck hitting the walls because they had previously expected the puck to get around the walls. I grouped these kinds of disturbances together as demonstrating "strategic tensions."

Strategic tensions also include the tension when two or more students proposed different strategies for the next step while the student who at the moment was holding the mouse could only at most take one suggestion at a time. The proposed multiple, different strategies (often involved arguing) was a disturbance to the activity. The on-site decision about whether to take one of the suggestions, and/or which one to take, illustrates a strategic tension.

Then I grouped the disturbances related to the social elements: division of labor, community, and/or rules (as shown at the bottom of the activity model in Figure 3). I grouped them together as disturbances demonstrating "social tensions." For example, some disturbances involved students complaining about a particular student taking too much time to play without making any progress. Other students argued back that each student should be respected and given enough time to play if he or she had an idea to try. This kind of disturbance resulted from the division of labor (only one student could play at a time), the classroom rules (students should respect each other), and the community (students were playing together as a group).

The third group of disturbances was about the tool element—disturbances related to students' use of the tool, the Electric Field Hockey's interface, toolbars, and/or representations such as electric field lines. These disturbances occurred when students had different opinions about specific uses of a feature in the game, such as whether to click a button in the menu to activate a function embedded in the game. I grouped these disturbances together as demonstrating "tool tensions."

All the disturbances and tensions, at the same time, were related to the subject element—the students. The conceptual tensions involve students' ideas of the electrostatic phenomena; the tool tensions involve students' use of the tool; and the social tensions involve the social interactions among students. At the same time, I did not find any disturbances that were specifically related to the subject element. Previous studies have identified tensions in relation to the subject element, such as students as a "passive recipient" vs. "engaged learner" (Barab, Barnett, Yamagata-Lynch, Squire, & Keating, 2002). In the present study, I did not find any disturbances that can demonstrate a tension in relation to the students about what kinds of learners they were in the Electric Field Hockey activity.

It would be hard to fully separate the disturbances across the three themes identified, because, as AT claims, all the elements in the activity are interrelated. In the E-Field Hockey activity, students' exploration of the simulated electrostatic phenomena (the object) was embedded in the E-Field Hockey (the tool). It would be hard to separate strategic tensions from tool tensions. Also, because the group of students shared one tool, and collectively worked to achieve the object, all the strategic tensions and tool tensions reflect some aspects of social tensions. Sometimes, a disturbance involved multiple elements and illustrated multiple tensions. I categorized these three types of tensions for analytical purposes: to capture different aspects of the activity tensions. Conceptually, the three types were interconnected.

After identifying the tensions, I examined the tensions and to see in what ways the emergence of and students' coping with the tensions contributed to students' exploration of the simulated electrostatic phenomena.

**Data Selection**

The group of students I taught in the summer program was randomly assigned to me. I examined the transcripts of classroom videos and screen-recording videos for the three days and marked all disturbances. I checked all the disturbances to identify the three kinds of tensions. I found that in the beginning of the activity, disturbances were frequent and illustrated all three types of tensions. This should not be surprising because, as Engeström (2001) pointed out, when an activity system adopts a new element, such as a new technology or a new object, it often leads to contradictions. Therefore, at the beginning of the activity when the E-Field Hockey was newly introduced, disturbances were many and tensions emerged. Overtime, the occurrence of the disturbances reduced, although they did not totally disappear.

When I present data in the results section, I will present detailed descriptions of the beginning of the activity, roughly the first six minutes of the activity, to illustrate the occurrence of disturbances and the formation of tensions. At the end of six minute the camera recording the front board stopped working. Thus these six minutes of the activity was the episode that (1) I had detailed data of; and (2) was in the beginning of the activity. I call these six minutes the beginning episode. After describing the beginning episode, I will show, by presenting illustrative transcripts from all the three days, how each type of tension developed during the activity, and how the tensions influenced students' exploration of the simulated electrostatic phenomena.

## Results

**Disturbances in the Beginning Episode**

This subsection describes the beginning episode, in which eight disturbances were marked.

**Disturbance 1.** I started the activity by projecting my computer screen, which was showing the interface of the E-Field Hockey's practice level, on the board. I introduced the rule of the E-Field Hockey interactive simulation: to make the positively charged, black hockey puck that starts on the left side of the field move and hit the goal that is on the right side of the field. In order to do this, the player needs to place the positive and negative balls on the field to attract and/or repel the hockey puck to move in a certain path. While I was talking, I dragged red balls and blue balls onto the field and then dragged them back to their boxes. I did not hit the start button to show how the puck would move accordingly. One student, Yujie, interrupted and asked: "Can you do us a demo?" I answered this question for the whole group: "You guys should try it by yourselves." One student agreed with me: "Right. It won't be fun if she (the teacher) does it." I marked here disturbance 1: students jumped in while I was introducing and spoke up with different opinions about who should do the first try.

I then asked students who would like to do the first try. One student took the mouse and started to place balls on the field. He put one red ball to the left of the hockey puck and then hit the start button (see Figure 4, left). The hockey puck shot off to the right and hit the goal. A *Ta-da...* sound came from the computer, and "Goal!" was displayed on the screen (see Figure 4, right).

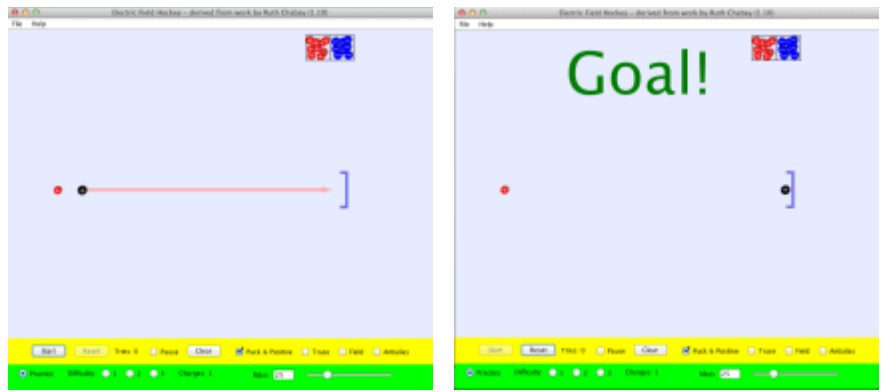

Figure 4. The first student's strategy (left) and result (right), reproduced according to the classroom video.

**Disturbance 2.** Then I asked if anybody else also wanted to try. Yujie, who had asked me to do a demo, took the mouse. Under my direction, he clicked the clear button to remove the red ball from the field, and then clicked the reset button to bring the hockey puck back to its starting position. He then dragged a red ball from the box and put it to the left of the hockey puck. Seeing this, another student, Yizhi, immediately commented, "This is just the same [as the first student's strategy]." And he repeatedly suggested, "Put a negative [charged ball] on the other end." "Use a negative one." Yujie did not heed Yizhi's suggestion but continued to put a second red ball closely under the first red ball (see Figure 5, left). Yizhi reacted, "What…? How could this work?" Yujie finished placing the balls and hit the start button. The hockey puck shot to the right and successfully hit the goal.

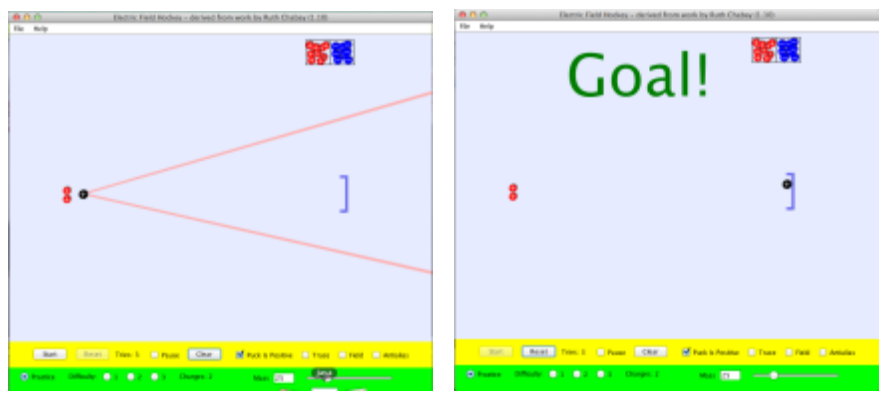

Figure 5. Yujie's strategy (left) and result (right), reproduced according to the classroom video.

I marked this episode disturbance 2: Yizhi was trying to impose his ideas on Yujie's strategy while Yujie was operating the E-Field Hockey, but was then surprised that Yujie's strategy worked: it made the hockey puck hit the goal.

**Disturbances 3-5.** Then the mouse was handed to a third student, Hanzhong. He cleared the field and reset the hockey puck to its starting position. He then dragged two blue balls and put them very close to each other left of (or inside) the goal (Figure 6, left). Yujie commented on it right away, "This way the puck won't get into the goal." Another student disagreed with Yujie: "Yes. It (the puck) could [get to the goal]."

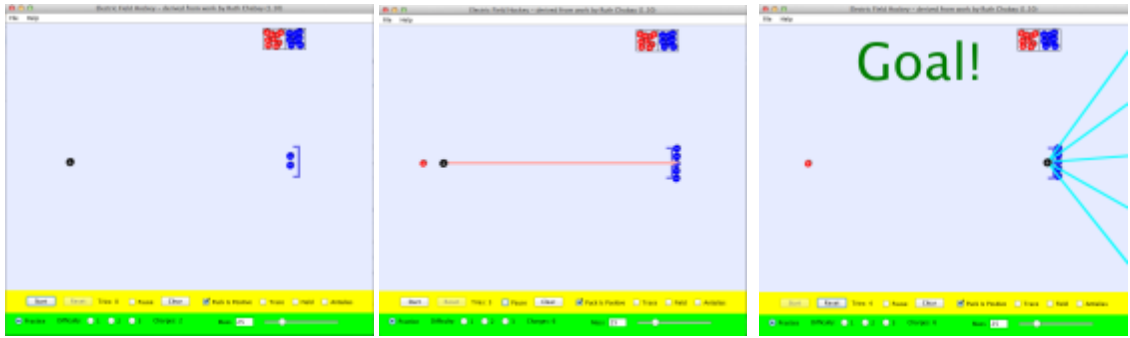

Figure 6. Left: Hanzhong put two blue balls on the left of the goal. Middle: Hanzhong's final strategy. Right: The hockey puck hit the goal. Reproduced according to the classroom video.

Hanzhong then moved the two blue balls to the right side (or outside) of the goal and then continued to line up more blue balls vertically (see Figure 6, middle). Students wowed at his strategy to put in and line up so many blue balls. Hanzhong then added one red ball to the left of the hockey puck and hit the start button. The puck successfully rushed to the right and hit the goal (see Figure 6, Right).

I marked in this episode three more disturbances. Disturbance 3 was when Yujie commented on Hanzhong's original strategy to put two blue balls on the left side of the goal while Hanzhong was placing balls on the field, questioning whether the hockey puck would pass the blue balls and go to the goal. Yujie said the puck would not get into the goal, whereas another student said it would. Canonically, both ideas could be right depending on how the charged balls were set up (or coded) in the game by the developers. If the balls were set up as rigid bodies of certain sizes, then a negative ball would stop (or bounce back) the positive puck when the puck hits it. If the balls were set as symbols of no size nor rigidity, then the puck would rush "through" the blue balls and keep moving in the same direction. The E-Field Hockey set up the balls as the latter: symbols. Hanzhong, who was playing the game, did not test what would happen to the hockey puck with the two blue balls inside of the goal, but seemed to take Yujie's suggestion to move the balls to the right (Figure 6, middle).

Disturbance 4 occurred when the whole class marveled at Hanzhong lining up five blue balls closely on the right side of the goal. Hanzhong's strategy seemed unexpected to many students, so the whole group reacted to it with a surprised exclamation.

While Hanzhong was lining up the five blue balls on the field, Yujie asked me, "Can we remove the lines (the arrow diagrams attached to the puck)?" I did not hear him, so he asked me again. I heard his question the second time and told him, "No. We cannot." I marked this moment when Yujie asked about whether the lines could be removed as disturbance 5. Yujie's question about removing the arrows gently disturbed the flow of activity because students in class were mostly not focusing on the lines. Even though Yujie's question did not re-direct the group's attention toward the arrow diagrams, I still marked this as a disturbance because he brought up a topic that most of the other students were not expecting or focused on.

The lines Yujie had asked about were the solid, red or blue arrow lines attached to the hockey puck (see Figures 4 to 6). These lines are called a free-body diagram in physics: the arrow lines show the force vectors of the forces acting on the puck. When the red and blue balls are put on the field, arrow lines attached to the hockey puck appear automatically. The arrow lines represent the electric forces acting on the puck. Each line represents the force of a specific ball. The color of a line matches the color of the ball exerting the force (a red arrow line represents the force of a red ball, and a blue arrow line represents the force from a blue ball). The length of each arrow represents the magnitude of the force; the arrowhead represents its direction. When the force's source (the ball) is far away, the force will be very weak and the arrow line corresponding to that force will be too short to see. The free-body diagram on the puck is not optional. As soon as one puts a red or blue ball on the field, the arrow shows up instantly. When one puts a ball back in the storage bin, the corresponding line disappears instantly. When one moves a ball in the field, the corresponding line changes its length and direction in real time, representing the changing electric force vector.

**Disturbance 6.** At this moment, I thought that students had explored enough of the practice level and the class should move on to Level 1. A vertical wall appeared in the middle of the field (see Figure 1, top right, or Figure 7). The student who had the mouse at the moment placed a red ball to the bottom left of the puck as shown in Figure 7.

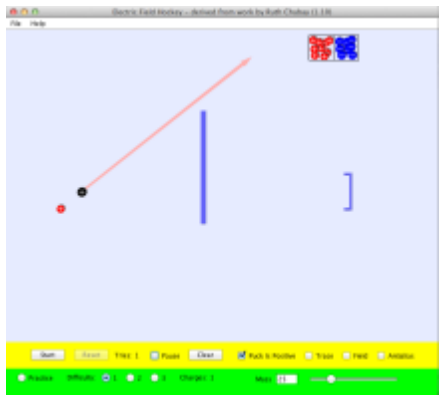

Figure 7. A student's first try at Level 1, which makes the hockey puck rush off

the field. Reproduced according to the classroom video.

Some students kept suggesting, "Add a blue ball." The student who was operating the mouse did not follow the suggestion and instead clicked the start button. The hockey puck quickly flew to the upper right and off the field. Not many students noticed the absence of the puck, so they continued to suggest adding balls to the field. One student noticed the puck was gone and alerted the others, "Hey! The puck disappeared!" The students were briefly silent and then burst into laughter, probably because they thought it was funny to "lose" the hockey puck and to realize that nobody even noticed that. I marked this as disturbance 6.

**Disturbance 7.** As the student who was operating the mouse experimented with strategies, Yizhi kept saying, "No. No. No…" claiming they would not work. Other students hushed him. Some students said, "Stop it. You should let him try." I marked this as disturbance 7: one student showed impatience toward the student operating the E-Field Hockey, and other students argued that each student should be respected and given opportunities.

**Disturbance 8.** At this point (five minutes and 38 seconds into the activity), the mouse was given to another student, Zixuan. He cleared the field, reset the puck, and then placed one red ball at the bottom left of the hockey puck's starting position, two red balls near the top of the wall, and two blue balls at the bottom right side of the goal, as shown in Figure 8 (left). He hit the start button. The puck moved to the right, collided with the wall, and stopped at the point of the collision. When the puck collided with the wall, "Collision!" appeared on the screen, and the computer emitted a colliding sound (see Figure 8, right).

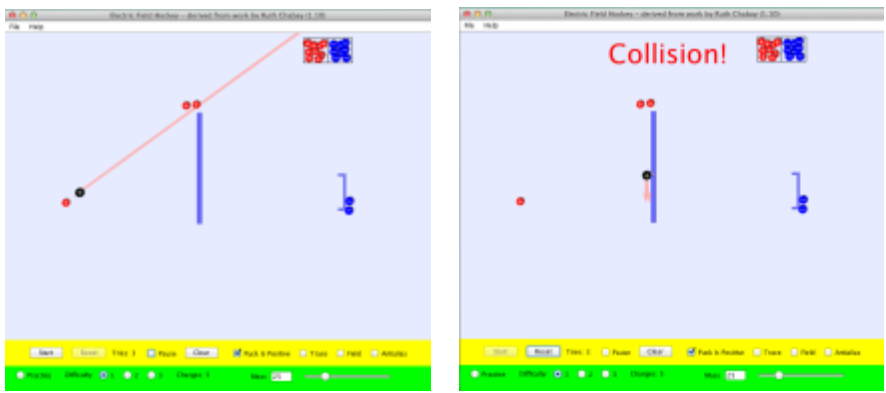

Figure 8. Left: Zixuan's strategy for Level 1. Right: The result of Zixuan's strategy. Reproduced according to the classroom video.

That was the first time students saw a collision in the E-Field Hockey, and they laughed at it. I identified this as disturbance 8. The somewhat funny colliding sound and the unexpectedness of it entertained the students and also brought their attention to the main challenge of the E-Field Hockey: to make the puck move in a curved trajectory—to move up and right, get around the wall, then move down and right, and then arrive at the goal.

These early disturbances described so far are summarized in Table 1.

Table 1. Disturbances during minutes 0-6 of Day 1.

| Time (Day 1) | Disturbances |
|---|---|
| 00:01:27 | 1. Yujie asked me to do a demonstration about the E-Field Hockey. |
| 00:02:26 | 2. Yizhi repeatedly suggested using a negatively charged ball when Yujie operated the mouse. |
| 00: 02:56 | 3. Yujie commented on Hanzhong's original strategy to put two blue balls left of the goal. |
| 00:03:11 | 4. The whole class marveled at how Hanzhong lined up five blue balls closely on the right side of the goal. |
| 00:03:40 | 5. Yujie asked about whether the arrow diagrams could be removed. |
| 00:04:27 | 6. Students laughed at the hockey puck rushing off the field. |
| 00:04:27 | 7. Some students stopped Yizhi from saying "No" to the student who was operating the mouse. |
| 00: 05:38 | 8. Students laughed when they saw the first collision. |

**Strategic Tensions**

I identified, from disturbances 2, 3, 4, 6, and 8 in the beginning episode, some "strategic tensions:" tensions about students' strategies of placing charged balls on the field. Sometimes, different students had different opinions about what kind of balls they wanted to use and about where to place the balls. Because the whole class shared one computer and only one student operated the mouse at a time, only one action could be taken as their very next step. When different strategies were proposed, the student who controlled the mouse needed to make a decision in-site and to take the action accordingly. This specific activity structure created some tensions when students had different strategies about their next steps and there were some observable disturbance related to these tensions. For example, disturbance 2 in the beginning episode illustrates a strategic tension. When Yujie, who at the moment controlled the mouse, put one red ball on the left of the hockey puck, Yizhi commented on that and said it was the same as the group's first try, and suggested to do something different, e.g., to put a blue ball on the right. In this case, Yizhi viewed Yujie as an agent of the group, rather than as an independent different individual, therefore should not repeat what had already been done by last person. Yujie did try a different strategy (two red balls on the left of the puck's starting position) on his turn, although it was not what Yizhi suggested (a blue ball on the right side of the hockey field). This vignettes showed that Yujie in part was similar with Yizhi in that he did not repeat a previous strategy. However, he also tried a different strategy that he was interested, not what someone else in the group suggested.

The specifics in each tension varied, although the tensions were all related to students' strategies. At the moment which disturbance 3 occurred, a tension arose because Yujie said that he did not believe that the puck would pass the blue balls that were placed before the goal and hit the goal, whereas some other students argued with him saying that it would. Other times, tensions arose because students saw phenomena that most of them did not expect, such as at disturbance 4 when Hanzhong lined up a group of blue balls on the right hand side of the goal to attract the hockey puck into the goal; at disturbance 6 when most students did not expect the puck to rush off the field and disappear; and at disturbance 8 when students saw the first collision.

At moments similar to disturbance 4, students marveled at Hanzhong's strategy but did not more action than the marveling, because although the strategy surprised the group, it worked to make the puck hit the goal. A tension existed briefly and students soon got over it. At moments similar to disturbances 6 and 8—a strategy did not work to make the puck hit the goal—students had to re-consider their current strategy and develop new strategies to make the puck hit the goal. The tension may keep for a longer period of time until students find better, working strategy. More disturbances similar to disturbances 6 and 8 repeatedly occurred during the rest of Day 1 and throughout Days 2 and 3. Because of the walls obstructing in the way between the puck's starting position and its goal, students frequently saw the puck rushing out of the field and/or the puck colliding with the wall(s) when they tried a strategy.

When students were experiencing strategic tensions, they decided to pick (e.g., between two proposed strategies) the strategy and/or to improve the previous unsuccessful strategy that can make the puck hit the goal. The students were then exploring the phenomena related to the strategies that they anticipated were at a better chance to make the puck hit the goal. In some occasions students gave up the opportunity to explore the phenomena that could be productive for the learning of the electric field just because those strategies appeared to have a smaller chance to make the puck hit the goal. For example, at disturbance 3, two students argued about whether the two blue balls placed before the goal would stop the puck from getting to the goal. Hanzhong, the student who at the moment held the mouse, did not try and see whether or not the puck would be stopped. Rather, he moved the two blue balls to the right of the goal and continued to add more blue balls along the same vertical line. It seems that he was focused on placing balls to get the greatest chance to make the puck hit the goal, rather than on exploring many plausible strategies.

**Social Tensions**

Disturbances 1 and 7 during the beginning episode demonstrated "social tensions:" tensions concerning the social structure of the activity. At the very beginning of the activity when disturbance 1 occurred, two students expressed different opinions about who should do the first demonstration: Yujie wanted to see me (the teacher) demonstrating whereas a different student wanted to try by themselves. This disturbance showed that students sitting in the same classroom, doing the same activity, held different views of the distribution of work, of the classroom rules, and of the classroom group that they were all parts of (corresponding to the elements of division of labor, rules, and community in the AT model shown in Figure 3). In disturbance 1, Yujie's request, asking the teacher to do a demonstration before students tried, showed that his understanding of the activity was more like a traditional classroom activity: teacher demonstrates the "right" way of operating the E-Field Hockey and students follow this way. The other student's voice, echoing the teacher that students should try by themselves, showed his understanding of the activity more of a student-led exploration that had less to do with finding right or wrong answers.

More social tensions emerged while the activity went on. These tensions had certain influences on students' exploration of the phenomena. For example, when Yizhi kept speaking "No." "No." loudly to multiple students who were operating the mouse, some of the other students hushed him and said that he should let the students try their strategies whether their strategies were better or not than the strategy he suggested (where I marked as disturbance 7). Explicitly supported by peers, the students who were operating the mouse continued to place the balls according to their strategies and explored some phenomena they wanted to see.

Social tensions occurred frequently during Day 1. When students were working on level 1, they continued to add, move, and remove the balls and try to make the puck get over the wall and to the goal. It was not easy. Students took turns with the mouse. They tried many strategies but the puck always hit the wall at different places. Some students started to complain that this was taking too much time. For example, following the beginning episode Zixuan tried many times but all failed. A student then asked if they could switch to a different person. Then Hanzhong took over the mouse and tried several times, but he, too, all failed. Yujie spoke up (at 8:31 minutes): "Change the person. This is wasting so much time." I responded to Yujie in front of the whole class, "Don't worry about the time, guys. Just take your time and try." During the seven minutes that followed, the mouse was passed from one student to the next. Several students tried, but all failed. At one point, Yujie got the mouse and played for a while but still did not make significant progress, either. A few students said something about his taking too much time. He immediately got defensive (at 16:19 minutes): "Actually I did not take much time…" I jumped in again and tried to put him at ease: "No worries. No worries. It doesn't matter if you take a long time..." Yujie continued acting defensively and pointed to other students who he thought took a longer time than he did: "Actually someone else did." Once again I tried to put the students at ease and stop their fighting. One student, Cai, echoed what I was saying (16:34 minutes): "Right. Not a big deal. It's just a game." The fight ended here. Cai took over the mouse (at 17:49 minutes), tried a few strategies, and finally made the puck arrive at the goal (at 18:30 minutes). That was the end of the first day of this activity.

During the first day students experienced frequent and heated arguments in relation to social tensions. During the second and third days disturbances in relation to social tensions decreased. Students seemed to reach a certain level of agreement in dealing with these tensions and focused more on the task.

When encountering social tensions, students' focus were not necessarily on completing the task. Rather, they started to learn to work as a group, to respect each other, and to achieve a shared goal while taking into account multiple points of view. Therefore, they were exploring phenomena not necessarily related to the most successful strategies, but were exploring phenomena more broadly.

**Tool Tensions**

Disturbances 3 and 5 in the beginning episode demonstrated "tool tensions." These tensions address students' use of the E-Field Hockey interactive simulation. Disturbance 3 has been discussed in the section of "strategic tensions:" Yujie interrupted Hanzhong when Hanzhong was placing two blue balls to the left of the goal. Yujie said that the puck would not pass the blue balls and hit the goal, whereas some other students argued that it would. Later on Day 1, one student, Cai, asked me two more questions about the balls. When she had just taken over the mouse and prepared to place balls, she asked me the same question that Yujie and Yizhi had argued about during disturbance 3, "Miss, I have question. Will the puck stop when it meets a blue ball?" I did not answer her directly but suggested she try and see. She then asked a second question, "Then are the balls in the bins unlimited? Like I can use as many as I want?" I nodded and said, "Yes."

The answer to this question depends on how the E-Field Hockey designers coded the puck and the balls in the E-Field Hockey. For example, the question about whether the hockey puck would pass blue balls depends on whether the balls were coded as rigid, finite-sized objects (then the puck will collide with the balls and bounce back) or symbols with no size and rigidity (then the puck will pass the balls and keep moving forward). Disturbance 3 could be about students' different predictions about a strategy (so it illustrates a strategic tension) and about their different views of the balls coded in the E-Field Hockey (so it also illustrates a tool tension). This disturbance was a good example of the fact that students' exploration of the simulated electrostatic phenomena was mediated by the tool, and that tensions about strategies and about the tool were entangled.

Disturbance 5 in the beginning episode was about the free-body diagrams embedded in the E-Field Hockey. Yujie noticed the free-body diagram and asked me whether the lines could be removed. His request to remove the diagram showed that the free body diagram might appear to Yujie as a distraction rather than a helpful representation. Looking more closely at the video of the beginning episode, I noticed that 2:26 minutes into the activity (the moment when disturbance 2 occurred, see Figure 5) when Yujie took the mouse and put a red ball to the left of the puck, he did not put the red ball in the same horizontal line with the puck and the goal, a red line appeared immediately and the line did not point to the goal. At this moment, Yizhi commented that this was just the same with the first strategy (one red ball was to the left of the puck). Then Yujie added the second red ball right below the first red ball (see Figure 5, left), another red line appeared and it did not point to the goal, either. Yizhi questioned on this strategy, "What? How could this work?" Yujie's strategy did make the puck hit the goal (see Figure 5, right), but before he hit the start button and before the puck actually arrived at the goal, at least for some students (such as Yizhi) appeared to doubt whether this strategy would make the puck hit the goal. It was possible that Yizhi's suspicion had to do with the fact that neither of the red lines was pointing to the goal. Then it was possible that Yujie viewed the lines rather distracting since he believed that his strategy would work. He would, then, rather remove these representations.

Unlike the strategic tensions and social tensions that emerged mostly during the first day (and then reduced in the following two days), the tool tensions were evident in both Days 1 and 2, and reduced during the last day. During the second day, students continued to explore level 2, which has two walls in the field. Two disturbances were identified that demonstrated tool tensions, shown in Table 3. The left column shows the times when the disturbances occurred. The middle column shows the recorded screens in relation to the disturbances. The right column describes the disturbances.

Table 3. Tool tensions on level 2.

| Time (Day 2) | Images on the screen | Strategy |
|---|---|---|
| 00:04:01 | 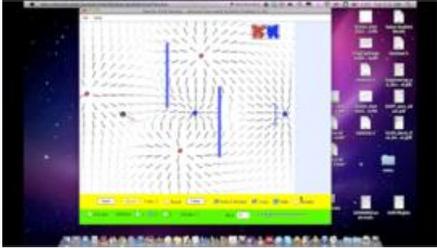 | **Try 7.** While students were playing on Try 7, Yizhi interrupted and suggested checking the three boxes in the yellow bar. The student playing the game did so. Then the little arrows (electric field line diagram) appeared on the screen. Most students did not expect these lines, and were amazed by them and said, "Wow!!" |
| 00:08:17 | 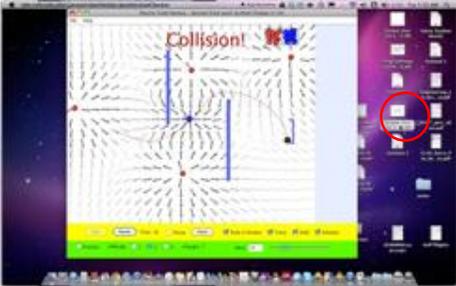 | **Try 16.** The strategy on Try 16 was not successful (the puck hit the edge of the goal). The playing student then tried to adjust and dragged a blue ball that was previously near the goal (not shown in the picture) to some other places. While he was dragging it, this student accidentally dragged the blue ball outside the interface and onto the computer desktop on the right hand side (in the red circle). Students in class did not expect this to happen (previously students moved all the balls within the interface) and were surprised and said, "Oops!" |

In the beginning of Day 2, students started with some strategies, but none of them made the puck get around the second wall. At try 7 (4:01 minutes), as the student operating the mouse was placing balls in the field and was about to hit the start button, Yizhi said, "Check the three [boxes]! At the bottom right [of the interface]. Those three [boxes]. Check them!" The student checked the three boxes per Yizhi's suggestion. The interface showed electric field lines (numerous little arrows spread across the field) on the screen (see the image in the fourth row). All the students in class expressed a collective "Wow" upon seeing it. Here I marked a disturbance. It illustrates a tension about whether and how to make use of the representations provided by the E-Field Hockey. Not many students expected these lines to appear by checking the boxes, nor did they know how to use them. Also on the previous day, one student (Yujie) had expressed his willingness to remove the free-body diagram. Yizhi, the student who suggested checking the boxes, later told me that he had tried this at home during the night before, so he knew what the lines would look like. That was why he suggested checking the boxes and making the representations visible. The electric field lines (numerous little arrows all over the field) visualize the electric field. The configuration of the field lines (thus the directions of the little arrows) changed in real-time when students added and moved the charges in the field. When the three boxes are checked, the puck's trajectory is also shown in a dashed red line after each try. The trajectory left behind the puck on the screen could allow students to see the path more clearly and spend more time reflecting on their strategies that led to the trajectory and then improve on their strategies.

Other than the exclamation upon seeing the representations for the first time, students did not talk explicitly about these representations. It was not clear whether and how they made use of the representations to develop and/or to improve strategies. However, there were some signs showing that students might have seen the representations helpful. For example, on Day 3, at the very beginning when students started with level 3, the operating student checked the three boxes before he placed any balls on the field. Nobody suggested him to do so, nor anybody reacted to him when he did it. It seemed that students had a consensus that they would like to see the representations while exploring level 3.

Another disturbance about tool tensions was shown in Table 3, the bottom row. At try 16 on level 2 the operating student was adjusting the position of the blue ball near the goal (not shown in the screenshot). He accidently dragged the blue ball out of the E-Field Hockey's interface and on to the desktop screen on the right hand side. The whole class went: "Oops!"

The disturbance concerned the design of the simulated hockey field area, the white background square enclosing the puck, the balls, the wall(s), and the goal. When the student dragged the blue ball out of the interface, that blue ball was no longer visible. However, when the playing student moved his mouse, the field lines in the field still changed accordingly, which indicated that the ball outside the interface was still affecting the electric field in the same way as the balls inside the interface did. The tension was resolved by seeing the changing field lines in the field. Students found that putting balls outside of the field still affects the hockey field, even though the balls are no longer visible.

Some students suggested dragging another blue ball from the bin into the field. Other students suggested to just give it a try as-is. The player took the latter suggestion and hit the start button. The puck still hit the lower edge of the goal. Then the player dragged another blue ball into the field and left of the goal. He failed a few more times and adjusted the blue ball's positions. At try 20 (10 minutes), the strategy finally made the puck successfully hit the goal!

The class broke into "Hooray!" to celebrate this success. Students spent less time to win level 2 (10 minutes) than they had spent on a supposedly easier level: level 1 (14 minutes).

During Day 3, at some tries, students purposely put balls outside of the simulated hockey field to improve their strategies (see in Table 4). The operating student dragged a blue ball from the storage box and put it on the right side (see the additional blue ball on the recorded screen of Try 48). At this try (Try 48), the hockey puck hit the second wall, but students seemed to acknowledge the effect of putting balls in the right zone. Multiple students continued to add, remove, and adjust positive and negative balls in this zone to try to make progress. Students spent 17 minutes and 52 tries on level 3 (during Day 3) but did not make the puck hit the goal before the class was dismissed. However, they made the puck get around three walls in spite of not hitting the goal, which was a very big achievement.

Table 4. Students purposely put balls outside of the simulated hockey field.

| Day Time | Image on the screen | Strategy |
|---|---|---|
| Day 3 00:16:09 | 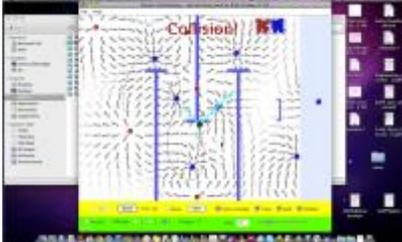 | **Level 3, Try 48.** The student purposely dragged another blue ball from the storage box and put it outside the hockey field (but still inside the game's interface). |

When encountering tool tensions, students were exploring the tool as well as the simulated electrostatic phenomena. Students' decision about what to do with a specific feature of the tool was largely oriented by their understanding of how that particular feature would help them develop a strategy to make the puck hit the goal. For example, when students accidentally found that they were allowed to place balls outside of the designated hockey field, they realized that they had more room to place balls. They then chose to place balls in those outside areas, even though it might violate the original rule of the tool. Students found that when the field lines and the trajectories were visible, they had more information about the electric field and about the hockey puck's motion, they chose to keep the representations visible so that they could improve their strategies based on the information. On the contrary, when a particular feature did not appear helpful for them to develop/improve their strategies (such as force diagrams), students would not be interested in using them.

## Discussion

**Mediated Experiences with Electrostatic Phenomena**

The goal of the E-Field Hockey activity was to allow students to explore electrostatic phenomena, which is hard to do with traditional lab equipment. Physics education researchers have suggested that if applied appropriately, a computer simulation can substitute the real-world laboratory to show physical phenomena and demonstrate physical laws that students are going to learn (Finkelstein et al., 2005). Some researchers and practitioners have investigated the effect of computer simulations in helping college students learn electromagnetism (Belcher & Olbert, 2003; Dori & Belcher, 2005). This paper adds to the literature, presenting a case study about how high school students explored the simulated electrostatic phenomena through an interactive computer simulation, the Electric Field Hockey. The E-Field Hockey interactive simulation highlighted electrostatic interactions and removed other real world factors such as air resistance. It allowed students to focus on exploring the simulated electrostatic phenomena (e.g., electric attraction and repulsion) and gave them some interactive experiences with simulated electric charges (e.g., to drag and move charges on the field).

In this activity, students' experiences with the electrostatic phenomena were mediated by the tool they used. The E-Field Hockey was designed based on a certain physical model. It is different from the real world phenomena. In the E-Field Hockey, the electric charges are depicted differently according to their charge type (red as positive and blue as negative). In reality, positive charges and negative charges do not have this color-coding. Similarly, the source charges are depicted as balls that will only exert forces but not experience forces from other charges. The balls were also designed to be immobile. The puck, on the other hand, represents the test charge that only experiences forces but does not exert forces on other charges, and can move. These modeled phenomena were different from the real world electrostatic phenomena. The E-Field Hockey added representations to visualize forces, field, and moving trajectory, which were invisible in the real world. When students encounter real world electrostatic phenomena, they will need to coordinate their experiences gained from the E-Field Hockey activity with what they might see differently in the real world.

**Tensions and Students' Exploration**

By looking for disturbances and examining them according to the activity model and the principle of contradictions, I identified three types of tensions in the E-Hockey activity: strategic tensions, social tensions, and tool tensions, illustrated in Figure 9. The strategic tensions address the object of the activity. The social tensions concern the rules, the community, and the division of labor in this activity. The tool tensions are about the E-Field Hockey interactive simulation. These tensions are categorized for analytical purposes but are practically intertwined.

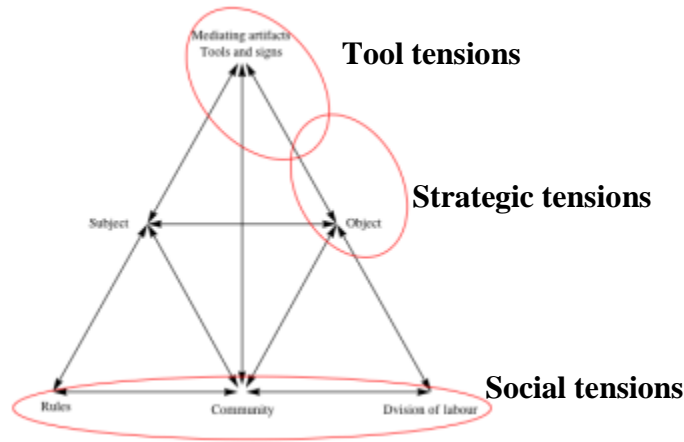

Figure 9. Strategic tensions, social tensions, and tool tensions

Engeström argues that when an activity system adopts a new element, such as a new technology or a new object, it often leads to a contradiction, and that such contradictions are innovative attempts to change the activity (Engeström, 2001). The introduction of the E-Field Hockey, the simulated electrostatic phenomena, and the organization of the activity are the stimuli that created tensions in the activity. It should not be surprising that in the beginning of the activity, when the E-Field Hockey was just introduced, disturbances and tensions occurred very often (e.g., eight disturbances during the first six minutes of the activity) and the types of tensions were many. Over time, the occurrence of the disturbances reduced to a certain level. For example, the social tensions occurred a lot during the first day, but during the second and third days, students managed the tensions and less and less disturbances were identified. The occurrence of the tool tensions shifted from exploring the features of the tool (e.g., whether the puck will stop at a blue ball; how many balls in the bin) to making use of the tool's features to develop better strategies (e.g., activate the representations to visualize the electric field and the charge's trajectory). The strategic tensions also occurred a lot at the beginning since that was when students first encountered the simulated electric charges and the main task in the E-Field

Hockey. Over time, when students became familiar with the E-Field Hockey, the strategic tensions decreased during the rest of the activity.

When students encountered strategic tensions, they usually prioritized exploring the phenomena that showed the hockey puck moving toward and hitting the goal, rather than exploring all the different phenomena. When students were experiencing social tensions they did not necessarily explore the phenomena that showed the puck hitting the goal, nor was their focus on the E-Field Hockey simulation. Rather, some students emphasized that they were a group and should respect each member even if a student were not making a promising strategy to make the puck hit the goal. When students encountered tool tensions, they explored the tool's features in relation to helping them develop/improve the strategies.

**Implications for Instruction and for Educational Technology Design**

The E-Field Hockey activity featured in this study was largely student-led. Besides introducing the E-Field Hockey and showing how to drag the blue and red balls on the field, the teacher did no more demonstration and/or instruction. The study examined how students achieved the object under such condition. In practice, the activity can be students and teacher co-led thus to make the exploration more productive. If a teacher wanted to scaffold students to explore the simulated phenomena in more productive ways, he or she should attend to these tensions and decide whether and how to intervene when tensions arise.

In the E-Field Hockey activity featured in the present study, sometimes the concerns of completing the task overrode the goal of exploring the simulated electrostatic phenomena. I suggest that these are the moments that a teacher should jump in and steer students to also explore those strategies that were productive for the learning of electrostatics even though they may not appear to have the highest chance to make the puck hit the goal. To help mitigate social tensions, the teacher can jump in when these tensions arise and acknowledge students' care about other students and their view of the group as a learning community. The teacher can also ask the operating students to explain his or her strategy to the group. The group, then, can listen to his or her ideas and students may shift to discuss the physics behind the strategies.

The key challenge—walls partially blocking the puck's path from the start to the goal—pushed students to operate the simulated electric charges in multiple, complex positions and observe the collective effects the charges have on the hockey puck. These features could help students explore the electrostatic phenomena in a way that traditional instruments often fail to do, e.g., the representations could help students visualize the electric field and the moving traces that are invisible in the physical world. At the same time, the E-Field Hockey could also be improved (e.g., make the free-body diagram optional; clarify how the balls are coded; etc.). Educators need to be aware that students' experiences with the electrostatic phenomena in this game are mediated by the tool and not the very same experiences they would have gained by interacting with the real world electric charged objects.

Cambridge, MA: Harvard University Press